# Modeling Complex Higher Order Patterns[*]

Zengyou He, Xiaofei Xu, Shengchun Deng

*Department of Computer Science and Engineering Harbin Institute of Technology,*

*92 West Dazhi Street, P.O Box 315, P. R. China, 150001*

Email: zengyouhe@yahoo.com, {xiaofei, dsc}@hit.edu.cn

**Abstract** The goal of this paper is to show that generalizing the notion of frequent patterns can be useful in extending association analysis to more complex higher order patterns. To that end, we describe a general framework for modeling a complex pattern based on evaluating the interestingness of its sub-patterns. A key goal of any framework is to allow people to more easily express, explore, and communicate ideas, and hence, we illustrate how our framework can be used to describe a variety of commonly used patterns, such as frequent patterns, frequent closed patterns, indirect association patterns, hub patterns and authority patterns. To further illustrate the usefulness of the framework, we also present two new kinds of patterns that derived from the framework: clique pattern and bi-clique pattern and illustrate their practical use.

## 1 Introduction

Frequent pattern (itemset) mining [1] is a widely studied topic in data mining research. Meanwhile, few efforts have been undertaken to extend frequent patterns to more complex and useful patterns such as frequent closed patterns [2], maximal frequent patterns [3] and indirect associations [4], etc. However, those efforts that do have been specific to the work at hand. Thus, an overall framework for understanding and extending frequent patterns is still lacking.

The goal of this paper is to provide such a framework and show its usefulness. Towards that end, this paper makes the following contributions:

**We introduce a framework for defining a pattern based on a view of constraint that is expressed as first order formulas on the interestingness of its sub-patterns**. Since a key goal of any framework is to allow people to more easily express, explore, and communicate ideas, we illustrate how our framework can be used to describe a variety of association patterns, from simple to complex. This includes traditional frequent patterns, as well as frequent closed patterns, maximal frequent patterns and indirect associations.

**We show that the framework provides us more insight on some traditional patterns to further enhance existing interestingness measures of association**. **To that end, we present the concept of sub-pattern interestingness curve to pictorially discriminate traditional defined patterns.** We illustrate this fact through examples based on frequent (closed) pattern and

[*] This work was supported by the High Technology Research and Development Program of China (No. 2002AA413310, No. 2003AA4Z2170, 2003AA413021) and the IBM SUR Research Fund.

statistical correlation pattern.

**We show that numerous new types of complex patterns could be designed under our framework.** More importantly, we present two new kinds of patterns that derived from the framework: clique pattern and bi-clique pattern and illustrate their practical use.

## 2 Traditional Frequent Patterns and Interestingness Measures

In this section, we first review the definitions of support-based concepts used in traditional transaction analysis. Agrawal formulated the problem of discovering frequent itemsets in market basket databases as follows [1]:

Let $I = \{i_1, i_2, \ldots, i_m\}$ be a set of $m$ literals called *items* and the database $D = \{t_1, t_2, \ldots, t_n\}$ a set of $n$ transactions, each consisting of a set of items from $I$. An itemset (pattern) $X$ is a non empty subset of $I$. The length of itemset $X$ is the number of items in $X$. An itemset of length $k$ is called a *k-itemset*. An itemset $Y$ is called a sub-itemset (or sub-pattern) of $X$ if $X \subseteq Y$ holds. A transaction $t \in D$ is said to contain itemset $X$ if $X \subseteq t$. The *support* of itemset $X$ is defined as the percentage of transactions in $D$ contain $X$, i.e., $support(X) = \| \{t \in D \mid X \subseteq t\} \| / \| \{t \in D\} \|$.

The problem of finding all *frequent itemsets* in $D$ is defined as follows. Given a user defined threshold *minisupport*, find all itemsets with supports greater or equal to *minisupport*. Frequent itemsets are also called *frequent patterns*.

An important property of support is the anti-monotone property: If $X$ and $Y$ are two itemsets where $X \subseteq Y$, then $support(Y) \leq support(X)$. The downward closure or anti-monotone property of support can be used to efficiently find frequent itemsets and is the foundation of the well-known Apriori algorithm [1].

To date, other interestingness measures such as interest factor, correlation, and entropy are also used to evaluate the interestingness of patterns- the stronger is the dependence relationship, the more interesting is the pattern. A more recent overview on other interestingness measures is provided in [5].

Regardless of how interestingness measures are defined, most previously studied patterns are evaluated using *only a single number* (interestingness value of the pattern), *without considering the interestingness of its sub-patterns*. A few exceptions among them are frequent closed patterns, maximal frequent patterns and indirect associations. However, an overall framework for modeling and unifying complex patterns is still lacking. To fulfill this void, in what follows, we shall focus on this issue.

## 3 A General Framework

### 3.1 Pattern Constraints and Representation

In this section, we describe a framework for understanding and modeling a pattern based on a view of constraint that is expressed as first order formulas on the interestingness of its sub-patterns. We then show how this framework can be used to express a variety of association patterns, including traditional frequent patterns, as well as frequent closed patterns, maximal frequent patterns and indirect associations.

For a given pattern (itemset) *X*, we let the power set of *X*, denoted by *Pow* (*X*), be the set of all unions of subsets of *X*. That is, we define *Pow* (*X*)= $\{S \mid S \subseteq X\}$. Hence, each sub-pattern of *X* is an element of *Pow* (*X*).

A *constraint C* is a predict on the set *Pow* (*X*). *Pow* (*X*) satisfies the constraint *C* if and only if *C* (*Pow* (*X*)) is true. We say that pattern *X* is *interesting* if and only if *C* (*Pow* (*X*))=true.

**Remarks**:

(1) Recent work has highlighted the importance of constraint-based mining paradigm in the context of association patterns [e.g., 6]. Some categories of constraints considered so far include item constraint, length constraint, model-based constraint and aggregate constraint. However, there are significant differences between our work and the previous researches in constrained frequent pattern mining with respect to both the types of constraints and research objective. Firstly, we mainly focus on constraints that are expressed as first order formulas on the interestingness of patterns. Secondly, our goal is to representing and constructing complex and meaningful patterns via a viewpoint of constraints on pattern's association.

(2) Another remark is about the interestingness measure used. Generally, interestingness measures involved in the constraint can be any previously defined measures such as support, confidence, interest factor, correlation, and entropy, etc.

In AI, machine learning and data mining, rules and graphs, including trees and networks, are popular representation schemas for patterns.

To depict a complex pattern that defined with constraint on the interestingness of its sub-patterns, we here utilize an *Item Hyper-Graph* (*IHG*) representation. *IHG* is a direct, simple and efficient representation for describing the constraint information at different sub-patterns

A hyper-graph *H* = (*V*, *E*) is a generalized graph, where *V* is a set of vertices and *E* is a set of hyper-edges. Each hyper-edge is a set of vertices that contains more than two vertices. In our model, each vertex $v \in V = X$ corresponds to an item in the pattern *X*, and each hyper-edge $e \in E$ is a sub-pattern that contained in *Pow* (*X*) and has a statistically significant association required by the constraint *C*.

Most statistical correlation measures don't posses the desired anti-monotone property. For example, the *lift* measure introduced by Brin et al. [7] is such kind measure. Hence, if we employ this kind measure for evaluating the interestingness of its sub-patterns, we could construct various constraints. For example, Fig. 1 shows some cases of different constraints on pattern {A,B,C}.The upper part of each case in this figure depicts the primary item occurrences and their pairwise statistical associations, while the lower part furnishes the hyper-graph representation of complex patterns that defined with constraint on the interestingness of sub-patterns. In Case 1, there exist statistically significant associations on two second order sub-patterns, [*A*, *B*] and [*B*, *C*]. Case 2 depicts a instance, such that statistically significant associations on all of the three second order sub-patterns, but not the [A,B,C] third order sub-pattern, exist. Case 3 shows a situation where association on the third order sub-pattern [*A*, *B*, *C*] exists, but there is no second order association between *A*, *B*, and *C*. Case 4 presents the state in which associations on the third order sub-pattern [*A*,*B* ,*C* ] and all of the three second order patterns [*A*,*B* ], [*B*,*C* ], and [*A*,*C* ] exist.

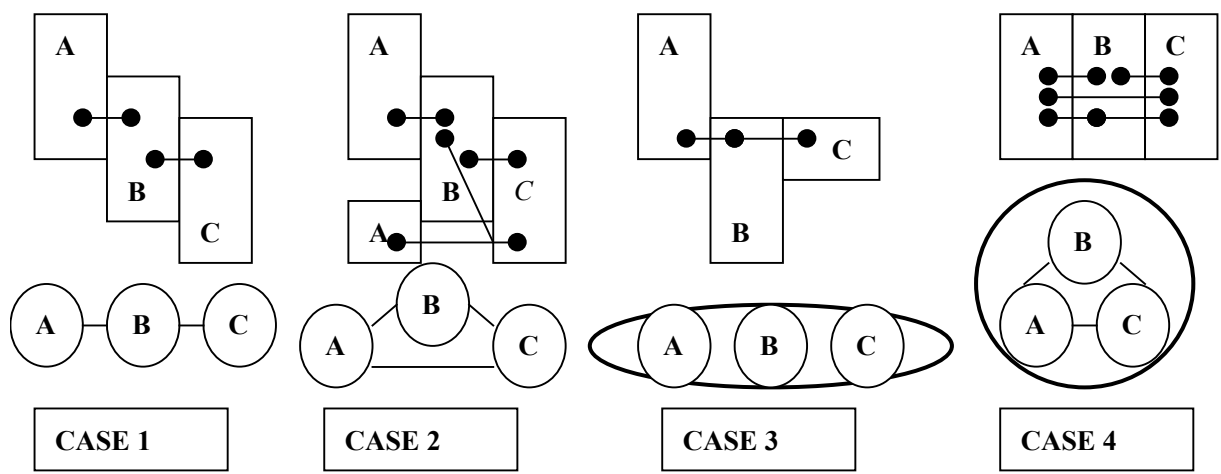

**Fig.1** Some cases of different constraints on pattern [A,B,C]

### 3.2 Example: Traditional Frequent Pattern

From the constraint-based viewpoint on patterns, we show how this framework can be used to express traditional frequent patterns in alternative ways.

A frequent pattern $X$ should satisfy the constraint: $support(X) \geq minisupport$. This constraint is very simple, since it only requires $X$ as an argument. Furthermore, due to the important anti-monotone property that is possessed by frequent pattern, an alternative constraint could be written as: $\forall S \in Pow(X), support(S) \geq minisupport$. In general, the latter constraint is stronger than the former one. For example, consider the cases of statistical correlation measure depicted in Fig.1. Hence, we argue that the concept of frequent pattern should be re-defined using the latter constraint on support values, although the original one is relatively simple.

### 3.2 Example: Frequent Closed Pattern

A frequent pattern $X$ is called *closed* if there exists no proper superset $Y \supset X$ with $support(X) = support(Y)$. From a constraint viewpoint, a *frequent closed pattern* $X$ satisfies:

$(support(X) \geq minisupport) \wedge (\neg \exists Y \supset X (support(Y) = support(X)))$

Previous researches have shown that frequent closed patterns are *lossless* in the sense that they uniquely determine the set of all frequent patterns and their *exact* frequency. At the same time frequent closed patterns can themselves be orders of magnitude smaller than all frequent patterns, especially on dense databases. Hence, a frequent closed pattern $X$ can contribute to the functionality of compression only and if only it satisfies the constraint: $\exists S (S \in Pow(X) \wedge S \neq X \wedge support(S) = support(X))$. More elements in $Pow(X)$ satisfy the constraint will result in larger compression ratio.

### 3.3 Example: Maximal Frequent Pattern

A frequent pattern $X$ is called *maximal* if it is not a subset of any other frequent patterns. In form of constraint, that is:

$$(\neg \exists Y \supset X(support(Y) \geq minisupport)) \wedge (support(X) \geq minisupport)$$

### 3.4 Example: Indirect Association Pattern

Indirect association is a more complex relation between itemsets than frequent pattern. Consider a pair of items *(a,b)* with a low support value, if there exists an pattern $M$ such that the presence of $a \cup M$ and $b \cup M$ are both frequent pattern, and *(a,b,M)* satisfy some certain dependency rule, then *(a,b)* are said to be indirectly associated via mediator-set $M$. Essentially, indirect association indicates a relationship of 'replacement' or 'competition' between items.

Formally, a pair of items *(a,b)* is said to be indirectly associated via a mediator set $M$ if the following conditions hold:

1. $support(\{a,b\}) < t_s$ (Itempair Support Condition).

2. There exists a non-empty set $M$ such that

(a). $support(\{a\} \cup M) \geq t_f$, $support(\{b\} \cup M) \geq t_f$ (Mediator Support Condition).

(b). $d(\{a\},M) \geq t_d$, $d(\{b\},M) \geq t_d$ where $d(P,Q)$ is a measure of the dependence between $P$ and $Q$ (Mediator Dependence Condition).

The above thresholds are called the itempair support threshold ($t_s$), dependence threshold ($t_f$) and frequent itemset threshold ($t_d$), respectively.

From a constraint viewpoint, an *indirect association pattern X* can be re-formulated as the pattern satisfies the following constraint:

$$\exists \{a\}, \{b\}, M(\{a\} \in Pow(X) \wedge \{b\} \in Pow(X) \wedge M \in Pow(X) \wedge (\{a\} \cup \{b\} \cup M = X)$$
$$support(\{a,b\}) < t_s \wedge support(\{a\} \cup M) \geq t_f \wedge support(\{b\} \cup M) \geq t_f \wedge d(\{a\},M) \geq t_d$$
$$\wedge d(\{b\},M) \geq t_d$$

With *Item Hyper-Graph* (*IHG*) representation, an indirect association pattern can be described as Fig.2 (the dashed line indicates a *negative* association), which is similar to the case 1 in Fig.1.

Apparently, it is trivial to extend indirect association between item pairs to indirect association between multiple items, as pictorially described in Fig.3. From Fig.3, we can see that it is more appropriate to call this kind of pattern as "star pattern". Besides indicating a relationship of 'replacement' or 'competition' between items, the "star pattern" also provides hints on "item ranking". That is, the center item (set) of the star is more important than other related items. Similar to the "star pattern", Wang et al. [8] studied hub patterns and authority patterns. Since star pattern is more general than hub and authority pattern, it is very natural to regard hub pattern and authority pattern as special cases of star patterns.

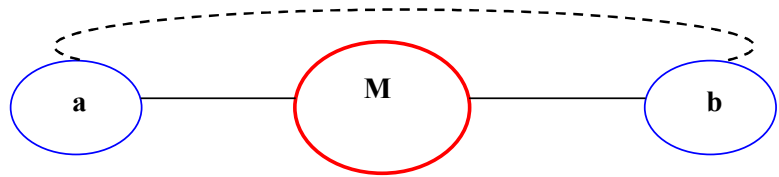

**Fig. 2** Graph representation of indirect association pattern

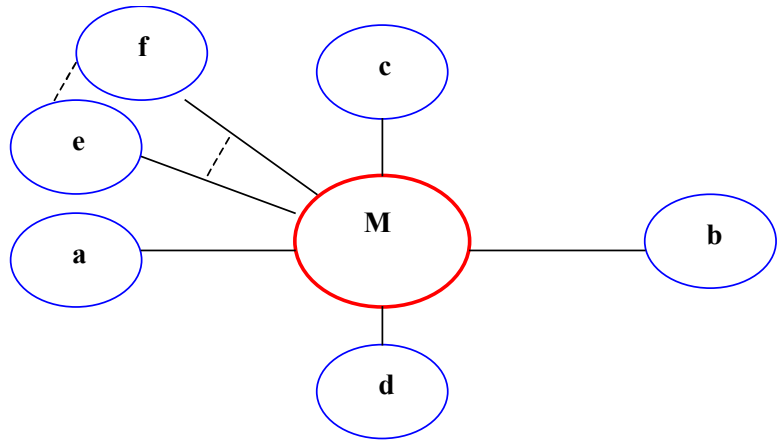

**Fig. 3** Graph representation of star pattern

# 4 More Insights on Traditional Patterns

In this Section, we show that the framework provides us more insight on some traditional patterns to further enhance existing interestingness measures of association. To that end, we present the concept of sub-pattern interestingness curve to pictorially discriminate traditional defined patterns. We illustrate this fact through examples based on frequent (closed) pattern and statistical correlation pattern.

### 4.1 Sub -Pattern Interestingness Curve

Suppose the elements in *Pow* (*X*) are ordered in a levelwise breath-first search manner. For example, an ordering for elements in *Pow* (*{A,B,C}*) is: *A, B, C, AB, AC, BC, ABC*. Since each sub-pattern of *X* has an *interestingness value*, we can graphically draw a curve to describe such a relationship (X-Axis: ordered elements in *Pow* (*X*); Y-Axis: interestingness values). This kind of curve is named as sub-pattern interestingness curve.

### 4.2 More insights on frequent pattern

With the help of sub-pattern interestingness curve (taking support as interestingness measure), we can get more insights on frequent patterns. For example, Fig.4 describes three different sub-pattern interestingness curves for *{A,B,C}*.

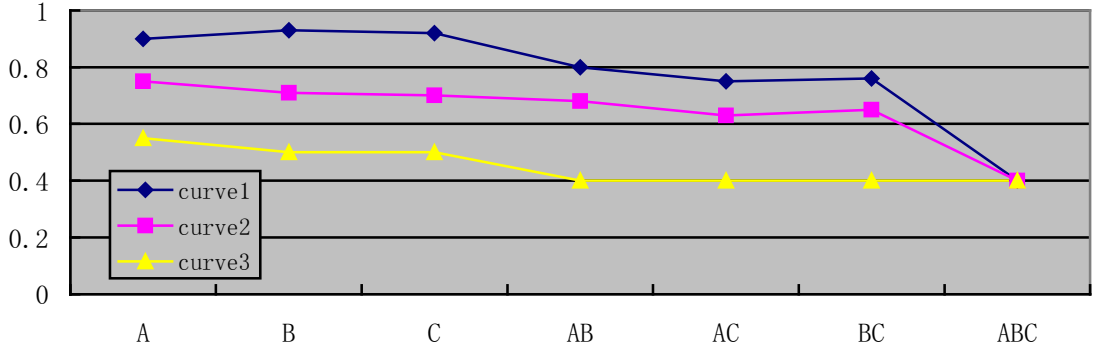

**Fig. 4** Three different sub-pattern interestingness curves for *{A,B,C}*(support)

In Fig.4, we can see that the support for *{A,B,C}* is the same (0.4) in all the curves. However, the distribution on the support values of sub-patterns of *{A,B,C}* are significantly different. Some important remarks are summarized as follows:

(1) All sub-patterns' supports exceed 0.4. That is, if we let 0.4 be the mini-support threshold, all three patterns described in curves are frequent and have the same support value. Furthermore, all curves are above the line of 0.4 (In the next section, we will get a different result on statistical measure based curves)

(2) From the traditional viewpoint of frequent pattern, all the three patterns will have the *same interestingness*, e.g., the same support value. However, with respect to the *interestingness* on sub-patterns, they are *definitely different*, since the fluctuation on the support values of sub-patterns varied significantly. Recently, Omiecinski [9] introduced two interesting measures, called *all confidence* and *bond*. Xiong et al. [10] independently proposed the *h-confidence* measure, which is equivalent to the *all confidence* measure of [9]. Both of measures introduced in [9] and [10] partially incorporated fluctuation on the support values of sub-patterns into the association measures.

(3) At last, we argue that traditional frequent patterns can further extended by considering the support values of sub-patterns, since the recent efforts in [9,10] towards this goal is still very limited.

### 4.2 More insights on statistical correlation pattern

Since most statistical correlation measures don't posses the desired anti-monotone property. Therefore, compared to the frequent pattern, the sub-pattern interestingness curves with statistical correlation measures are more complex. For example, Fig.5 describes three different sub-pattern interestingness curves for *{A,B,C}*.

Similarly, if we only consider association on *{A,B,C}*, the patterns of different curves are equally interesting to the end user. However, from Fig.5, we can see that they are actually different kinds of patterns.

Firstly, different from frequent patterns, no guarantee is provided on the minimal correlation values of sub-patterns. Apparently, lack of anti-monotone property is the cause.

Secondly, incorporating the sub-pattern correlation, we present two new kinds of patterns and argue that they will be interesting: *all-correlation* pattern and *unexpected-correlation* pattern.

Our formulations on the new patterns are as following (*col* (*S*) represents the statistical correlation of pattern *S*; *min_correlation* is the minimal correlation threshold).

(1) *All-correlation* pattern

A pattern *X* is called *all-correlation* pattern if and if only it satisfies the constraint:

$$\forall S \in Pow(X), col(S) \geq mini\_correlation$$

(2) *Unexpected-correlation* pattern

A pattern *X* is called *unexpected -correlation* pattern if and if only it satisfies the constraint:

$$\forall S \in Pow(X)(col(S) < mini\_correlation \land S \neq X) \land col(X) \geq mini\_correlation$$

The definition of *all-correlation* pattern is motivated by the frequent pattern concept. Similar to frequent pattern, *all-correlation* pattern will have the desired anti-monotone property. Hence, unlikely most statistically defined patterns, *all-correlation* pattern can be discovered efficiently.

The definition of *unexpected -correlation* pattern is motivated by the fact that this kind of association is more likely to be suspicious, and needs to be checked by the user.

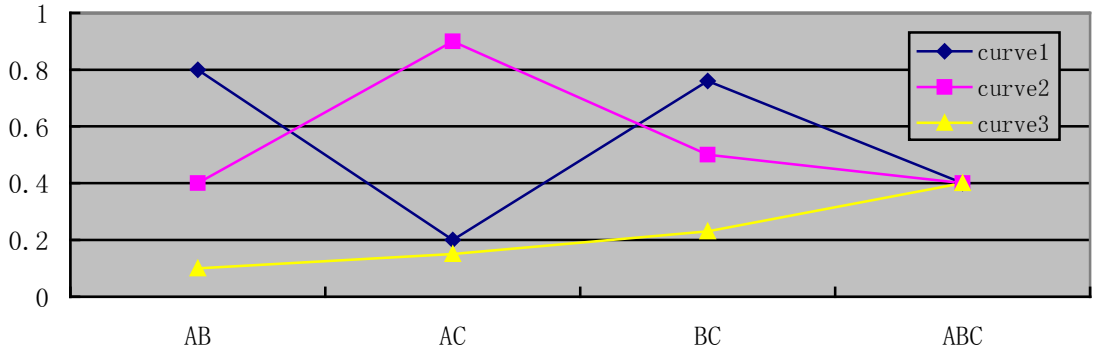

**Fig. 5** Three different sub-pattern interestingness curves for *{A,B,C}*( statistical correlation)

Finally, it should be noted other complex patterns could also be constructed under the framework.

# 5 Other Complex Patterns

We show that numerous new types of complex patterns could be designed under our framework. More importantly, we present two new kinds of patterns that derived from the framework: clique pattern and bi-clique pattern and illustrate their practical use.

### 5.1A Note on the number of potential complex patterns

Complex pattern will incur a huge construction space even without considering the interestingness measure. Since the size of *Pow* (*X*) is $2^{|X|}$, and each element in *Pow* (*X*) has two choice: satisfy the constraint or not. Hence, the number of potential types of patterns is at least $2^{2^{|X|}}$. This provides us great opportunities to construct useful and complex patterns in real applications.

In the next two sub-sections, we present two new kinds of patterns: clique pattern and bi-clique pattern and illustrate their practical use. Without loss of generality, support is used as the interestingness measure. Extension to other association measures is trivial.

## 5.2 Clique pattern

Motivated by the concept of clique from graph theory, a pattern *X* is called *clique* pattern if and if only it satisfies the constraint:

$$\forall S \in Pow(X)(|S| = 2 \rightarrow support(S) \geq minisupport)$$

That is, each sub-pattern of length 2 in *Pow* (*X*) is frequent. A sample clique pattern is depicted with *Item Hyper-Graph* in Fig.6.

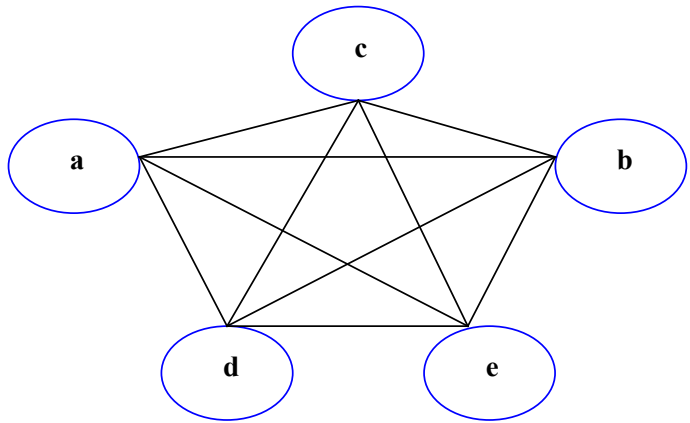

**Fig. 6** A sample clique pattern

According to the definition of clique pattern, we know that frequent pattern is a special case of clique pattern. That is, if *X* is a frequent pattern, it must be a clique pattern. This property has also been observed in previous research [11]. However, [11] aims to improve the efficiency of association mining, while our objective is to formally define clique pattern as a useful association pattern in real applications.

In real business applications, relationships between item pairs are easier to be understood and deployed. Hence, in most cases, knowing the underlying associations between all item pairs and organizing them into different cliques are more desirable.

Furthermore, without specific knowledge about the target data, users will have difficulties in setting the support threshold to obtain their required results in frequent pattern mining. If the support threshold is set too large, there may be only a small number of results or even no result. In which case, the user may have to guess a smaller threshold and do the mining again, which may or may not give a better result. If the threshold is too small, there may be too many results for the users; too many results can imply an exceedingly long time in the computation, and also extra efforts to screen the answers. Therefore, discovering the more general clique patterns will be a good alternative for solve above-mentioned problem in frequent pattern mining.

## 5.3 Bi-Clique pattern

A pattern *X* is called *bi-clique* pattern if and if only it satisfies the constraint:

$$\exists W, V \in Pow(X)(W \cup V = X \wedge W \cap V = \Phi \wedge$$
$$(\forall w \in W, v \in V(support(\{w,v\}) \geq minisupport)) \wedge (\forall a, b \in W(support(\{a,b\}) < minisupport))$$
$$(\forall c, d \in v(support(\{c,d\}) < minisupport))$$

That is, there exists two disjoint sub-patterns $W$, $V$ in $Pow$ ($X$), each item pair is frequent if and only if they are not in the same sub-set. A sample bi-clique pattern is depicted with *Item Hyper-Graph* in Fig.7.

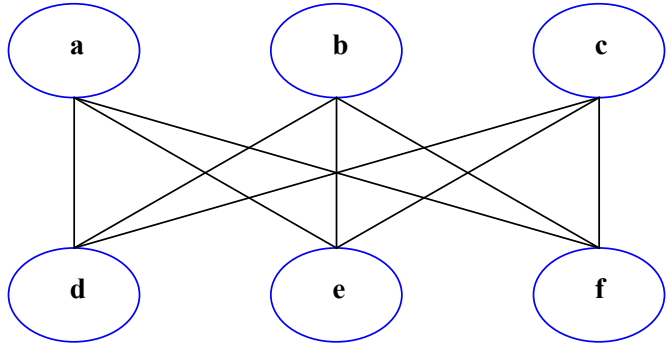

**Fig. 7** A sample bi-clique pattern

A bi-clique pattern describes the situation of cross association between two itemset. Clearly, such pattern not only indicates a more complex relationship of 'replacement' or 'competition' between items than indirect association, but also presents two incompatible clusters of items. Undoubtedly, such kinds of patterns have many potential applications in marketing and product management.

## 6 Conclusions

In this paper, we describe a general constraint-based framework for modeling a complex pattern based on evaluating the interestingness of its sub-patterns. Our framework can be used to describe a variety of commonly used patterns, such as frequent patterns, frequent closed patterns, indirect association patterns, hub patterns and authority patterns. Consequently, we also present two new kinds of patterns that derived from the framework: clique pattern and bi-clique pattern and illustrate their practical use.

In the future, we are planning to construct other complex but meaningful patterns under the framework. At the same time, efficient algorithms for mining those patterns will also be addressed.